\documentclass[10pt,final,journal]{IEEEtranTCOM}

\normalsize

\usepackage[utf8]{inputenc}
\usepackage{cite}
\usepackage{amsmath}
\usepackage{amsfonts}
\usepackage{acronym}
\usepackage{cleveref}
\usepackage{algorithmic}
\usepackage{algorithm}

\ifCLASSINFOpdf
\usepackage[pdftex]{graphicx}
\else
\usepackage[dvips]{graphicx}
\fi

\newacro{MMSE}{Minimum Mean Squared Error}
\newacro{MSE}{Mean Squared Error}
\newacro{MAC}{Multiple Access Channel}
\newacro{SQLC}{Scalar Quantizer Linear Coding}
\newacro{DQLC}{Distributed Quantizer Linear Coding}
\newacro{JSCC}{Joint Source-Channel Coding}
\newacro{AWGN}{Additive White Gaussian Noise}
\newacro{KF}{Kalman Filter}
\newacro{pdf}{probability density function}
\newacro{OPTA}{Optimal Performance Theoretically Attainable}
\newacro{SDR}{Signal-to-Distortion Ratio}
\newacro{SNR}{Signal-to-Noise Ratio}
\newacro{BC}{Broadcast Channel}
\newacro{WSN}{Wireless Sensor Network}
\newacro{MIMO}{Multiple-Input Multiple-Output}

\def\B{\boldsymbol}

\newtheorem{lemma}{Lemma}

\usepackage{tikz}
\usetikzlibrary{calc}

\newcommand{\positiontextbox}[4][]{%
	\begin{tikzpicture}[remember picture,overlay]
		\node[inner sep=3pt, fill=yellow,align=left,draw,line width=1pt,#1] at ($(current page.north west) + (#2,-#3)$) {\parbox{.95\paperwidth}{#4}};
	\end{tikzpicture}%
}

\begin{document}

	\onecolumn
	\begingroup
	
	\setlength\parindent{0pt}
	\fontsize{14}{14}\selectfont
	
	\vspace{1cm} 
	\textbf{This is an ACCEPTED VERSION of the following published document:}
	
	\vspace{1cm} 
	O. Fresnedo, P. Suárez-Casal and L. Castedo,``Transmission of Spatio-Temporal Correlated Sources over Fading Multiple Access Channels with DQLC Mappings'', \textit{IEEE Transactions on Communications}, vol. 67, n.o 8, pp. 5604-5617, Aug 2019, doi: 10.1109/TCOMM.2019.2912571
	
	\vspace{1cm} 
	Link to published version: https://doi.org/10.1109/TCOMM.2019.2912571
	
	\vspace{3cm}
	
	\textbf{General rights:}
	
	\vspace{1cm} 
	\textcopyright 2019 IEEE. This version of the article has been accepted for publication, after peer review. {Personal use of this material is permitted. Permission from IEEE must be obtained for all other uses, in any current or future media, including reprinting/republishing this material for advertising or promotional purposes, creating new collective works, for resale or redistribution to servers or lists, or reuse of any copyrighted component of this work in other works.}
	\twocolumn
	\endgroup
	\clearpage
\title{Transmission of Spatio-Temporal Correlated Sources over Fading Multiple Access Channels with DQLC Mappings}

\author{Óscar~Fresnedo~\IEEEmembership{Member,~IEEE},
	Pedro~Suárez-Casal,
	Luis~Castedo,~\IEEEmembership{Senior Member,~IEEE}
	\thanks{Óscar Fresnedo, Pedro Suárez-Casal and Luis Castedo are with the Department
		of Computer Engineering, Universidade da Coruña, 15071, A Coruña, Spain,
		e-mail: \{oscar.fresnedo, pedro.scasal, luis\}@udc.es.}%
}

\maketitle

\begin{abstract}
The design of zero-delay \ac{JSCC} schemes for the transmission of correlated information over fading \acp{MAC} is an interesting problem for many communication scenarios like \acp{WSN}.
Among the different \ac{JSCC} schemes so far proposed for this scenario, \ac{DQLC} represents an appealing solution since it is able to outperform uncoded transmissions for any correlation level at high \acp{SNR} with a low computational cost. In this work, we extend the design of \ac{DQLC}-based schemes for fading \acp{MAC} considering sphere decoding to make the optimal \ac{MMSE} estimation computationally affordable for an arbitrary number of transmit users. The use of sphere decoding also allows to formulate a practical algorithm for the optimization of \ac{DQLC}-based systems.
Finally, non-linear Kalman Filtering for \ac{DQLC} is considered to jointly exploit the temporal and spatial correlation of the source symbols. Results of computer experiments show that the proposed \ac{DQLC} scheme with the Kalman Filter decoding approach clearly outperforms uncoded transmissions for medium and high \acp{SNR}. 

\end{abstract}


\IEEEpeerreviewmaketitle

\acresetall

\positiontextbox{11cm}{27cm}{\footnotesize \textcopyright 2019 IEEE. This version of the article has been accepted for publication, after peer review. Personal use of this material is permitted. Permission from IEEE must be obtained for all other uses, in any current or future media, including reprinting/republishing this material for advertising or promotional purposes, creating new collective works, for resale or redistribution to servers or lists, or reuse of any copyrighted component of this work in other works. Published version:
	https://doi.org/10.1109/TCOMM.2019.2912571}

\section{Introduction}

\IEEEPARstart{I}{n} this paper we address the design of zero-delay analog \ac{JSCC} schemes for the distributed transmission of correlated analog information from several devices to one central receiver over a common wireless channel, i.e., a fading \ac{MAC}. This is an interesting problem useful to model a large number of communication scenarios like, for example, several sensors periodically sending their data to a central node in a \ac{WSN}. Measurements from sensors are typically correlated both in time and space, and this information must be taken into consideration in the design of \acp{WSN}.

Existing approaches for transmission over a fading MAC mostly lie on source-channel coding separation. %
In this case, some sort of distributed source encoding scheme is used to exploit the source correlation \cite{slepian73,Wyner76} and its output is next encoded with a capacity-achieving channel encoder. 
Transmission schemes based on source-channel separation, however, require large block sizes at both encoders, and hence they are not particularly adequate for an intermittent transmission of small amounts of data. In addition, source-channel separation is not always optimal for multiuser communications, especially when the information among users is correlated \cite{cover80,gastpar08,tian13}.
 
As an alternative, several works in the literature advocate the use of \ac{JSCC}, replacing the two separated source and channel encoders by a single joint encoder which maps the correlated source information to the corresponding channel symbols. In \cite{lapidoth10}, a non-linear hybrid \ac{JSCC} scheme is proposed for the transmission of bivariate Gaussian sources. This scheme, which combines vector quantization and uncoded transmission, is able to outperform any system based on source-channel separation and provides the minimum achievable distortion in those \ac{SNR} ranges where the uncoded scheme is no longer optimal. However, a practical implementation of vector quantization also requires large block sizes and a high computational cost to achieve a performance close to the theoretical bounds. A practical \ac{JSCC} scheme based on vector quantization with arbitrary block sizes is presented in \cite{floor15} for multivariate Gaussian sources. In \cite{floor15}, the authors also propose a zero-delay \ac{JSCC} mapping, named \ac{DQLC}, helpful for applications with strict delay constraints. This scheme is based on quantizing the symbols of all users except those of one user that are simply scaled and truncated prior to its transmission. A formal derivation of the optimal zero-delay \ac{JSCC} mapping for the case of bivariate Gaussian sources is carried out in \cite{kron14}, where a non-parametric version of the \ac{DQLC} was obtained. Other non-linear \ac{JSCC} schemes have been considered for different scenarios like mappings based on lattice coding for the orthogonal transmission of correlated information in Gaussian \acp{MAC} \cite{Karlsson11} or channel-optimized vector quantization for Gaussian \ac{MIMO} \acp{BC} \cite{Persson12}. %
 
In this paper, we address the zero-delay distributed \ac{JSCC} of spatially and temporally correlated sources in the \ac{MAC} considering an arbitrary number of users and fading channels. On one side, spatial correlation is exploited by using a \ac{DQLC} scheme which is conveniently optimized depending on the channel conditions. At the receiver, an estimate of the user symbols is jointly computed with the \ac{MMSE} decoder. In addition, a sphere decoder is considered to lower the computational cost of the decoding operation. Sphere decoding is a general strategy to search for the closest vectors in a lattice and it was originally proposed to detect digital signals in \ac{MIMO} transmissions \cite{Hochwald03}. This strategy has already been applied to lower the computational cost of the \ac{MMSE} estimation when transmitting analog symbols using modulo mappings \cite{Suarez17}. In this paper, we mathematically derive the lattice corresponding to \ac{DQLC}-based mappings and apply the sphere decoder to the obtained lattice. On the other side, a non-linear version of the \ac{KF} for \ac{DQLC} is proposed to leverage the temporal correlation in the source vectors at two consecutive time instants. \ac{KF} is a well known algorithm based on linear equations to estimate parameters observed along time \cite{kalman1960new}. It has been previously applied to communication problems such as channel tracking \cite{Zhiqiang02} or state estimation of \acp{WSN} with distributed source coding \cite{Quevedo10}. The use of non-linear \ac{KF} techniques to exploit the temporal correlation in the decoding of analog \ac{JSCC} symbols has been also applied to modulo-like mappings \cite{Suarez18}. In this paper, we follow a similar approach based on the idea of computing a prediction with the statistical information corresponding to the transition phase and using the received symbols to refine such a prediction. As in the case of sphere decoding, the integration of \ac{KF} with \ac{DQLC} mappings produces different mathematical expressions which lead to a specific development for the computation of the overall lattice.

\subsection{Contributions}

The main contributions of this work are the following:
\begin{itemize}
	\item A feasible implementation of the optimal \ac{MMSE} decoder for the \ac{DQLC} mapping in scenarios with an arbitrary number of users and fading. The proposed decoder relies on the use of a sphere decoder to lower the overall complexity while computing the numerical integrals involved in the \ac{MMSE} decoding. Although the approach resembles that in \cite{Suarez17}, the transformations over the original mapping function and the mathematical derivation of the searching lattice are different.  
	\item A practical optimization of the \ac{DQLC} mapping parameters which enables its utilization on scenarios with a moderate number of transmit users. 
	\item The integration of Kalman Filtering techniques into the communication scheme based on \ac{DQLC} to exploit the temporal correlation of the source information.
\end{itemize}

\section{System Model} \label{sec:model}

\begin{figure}[!t]
	\centering
	\includegraphics[width=0.85\columnwidth]{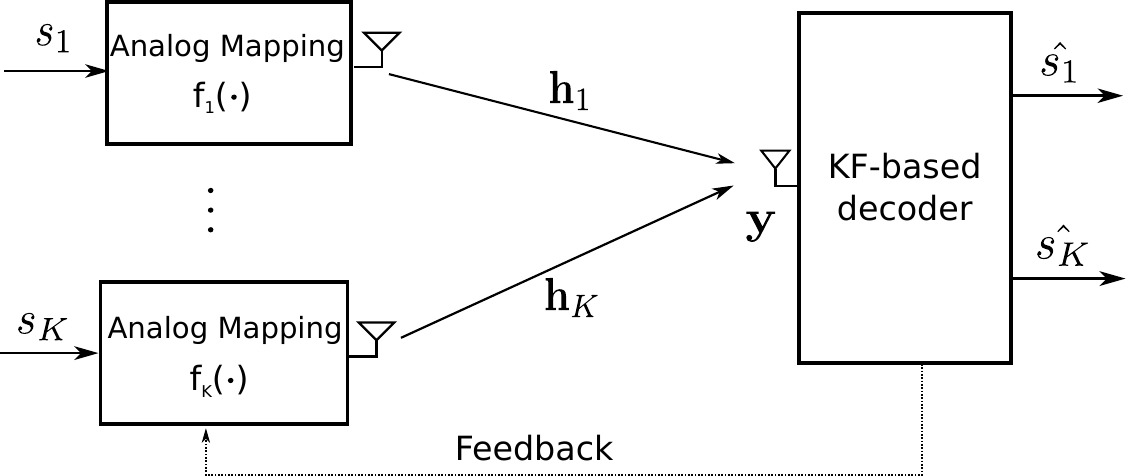}
	\caption{Block diagram of the considered \ac{MAC} scenario.}
	\label{fig:system_model}
\end{figure}

Let us consider the fading \ac{MAC} model shown in \Cref{fig:system_model}, where $K$ single-antenna users transmit their information to a central node, also equipped with a single antenna, over a wireless channel. The source information is assumed to be correlated both spatially and temporally according to the following autoregressive model
\begin{align}
\B{s}_{t} = \B{F}\B{s}_{t-1} + \B{w}_t,~~\forall t = 1,2, \ldots T
\label{eq:transition_eq}
\end{align}
where $\B{s}_{t} \in \mathbb{C}^{K\times1}$ represents the vector of user symbols at the $t$-th time instant, $\B{F} \in \mathbb{C}^{K\times K}$ is a diagonal matrix which models the state transitions, and $\B{w}_t \sim \mathcal{N}_{\mathbb{C}}(\B{0}, \B{C}_w)$ is the process noise component with covariance $\B{C}_w$. The vector of $K$ user symbols at each time instant  is assumed to follow a multivariate circularly symmetric complex-valued Gaussian distribution with zero mean
and covariance matrix $\mathbf{C}_{\B{s}}$, i.e.,  $\B{s}_{t}\sim \mathcal{N}(\B{0}, \B{C}_{\B{s}})$. Without loss of generality, we assume that $[\B{C}_{\B{s}}]_{k,k}=1 ~\forall k$, while $[\B{C}_{\B{s}}]_{i,j} = \rho_{i,j},~ i\neq j$, represents the correlation between the source symbols corresponding to the $i$-th and $j$-th users.  

At each time instant $t$, the source symbols are individually encoded at each user using a particular zero-delay analog \ac{JSCC} mapping $\B{f}(\cdot): \mathbb{C}^K \rightarrow \mathbb{C}^K$, which maps the vector of $K$ source symbols into the corresponding $K$ channel symbols. In this work, we focus on \ac{DQLC} since it has been shown to provide good performance for the considered \ac{MAC} scenario with \ac{AWGN} channels \cite{floor15}. 
The resulting symbols are then transmitted over the MAC and hence the received signal is
\begin{align}
y_t = \B{h}_t^T\B{f}(\B{s}_t) + n_t,
\label{eq:observation_eq}
\end{align}
where $\B{h}_t = [h_{t,1},  \ldots, h_{t,K}]^T$ represents the fading \ac{MAC} response, 
$n_t \sim \mathcal{N}_{\mathbb{C}}(0, \sigma_n^2)$ is the \ac{AWGN} component, and the distributed mapping function can be represented in a vector form as $\B{f}(\B{s}_t)=[f_1(s_{t,1}), \ldots, f_K(s_{t,K})]^T$. Finally, per-user individual power constraints are assumed in the form of $E\left[|f_k(s_{t,k})|^2\right] \le T_k$. Note that Eq. \eqref{eq:transition_eq} and \eqref{eq:observation_eq} constitute the transition and the observation steps, respectively,  of a non-linear \ac{KF} setting.

The user channels are assumed to be perfectly known at the receiver. This information is employed to determine the optimal values for the parameters of the mapping function $\B{f}(\cdot)$. Those optimal parameters are then sent to the corresponding users over a noiseless feedback channel.
At the receiver, an estimate of the source symbols is also computed from the received symbol $y_t$, using the channel information. Since we consider the transmission of continuous-amplitude information, the objective of the communication system is to minimize the distortion between the source and estimated symbols according to the \ac{MSE} criterion. The distortion between the source and decoded symbols is hence determined as
\begin{align}
\xi = \frac1{KT}\sum_{t=1}^T\sum_{k=1}^K |s_{t,k} - \hat{s}_{t,k}|^2,
\label{eq:sum_mse}
\end{align}
where $\hat{s}_{t,k}$ is the estimate of the $k$-th source symbol at the time instant $t$.

In such a case, the optimal decoding is the one that minimizes the average \ac{MSE} between both pair of vectors. However, when the mapping functions are non-linear, \ac{MMSE} decoding usually requires to numerically solve the corresponding integrals using Monte Carlo methods, which significantly increases the overall computational cost, especially when the number of dimensions grows. For this reason, a practical implementation of those mappings requires the design of decoding strategies with an affordable complexity.

The system variables corresponding to the source symbols, user channels, and noise components are complex-valued with uncorrelated real and imaginary parts. Hence, the system model presented in the previous section can be transformed into an equivalent real-valued one to simplify the notation \cite{Suarez17}. This transformation just implies that the dimension of the different variables and functions is doubled. Henceforth, the subindex $t$ is disregarded also for simplicity.

In the ensuing sections, we will describe the \ac{DQLC}-based mappings proposed to encode the user information and its corresponding \ac{MMSE} decoder.

\section{DQLC scheme}\label{sec:dqlc}

As introduced in \Cref{sec:model}, DQLC will be employed to encode the source symbols into the corresponding channel symbols. DQLC is a distributed \ac{JSCC} mapping function proposed to transmit multivariate Gaussian sources over a Gaussian MAC \cite{floor15}.  Mathematically, this mapping function is given by 
\begin{align}
f_k(s_{k})=\left\{
\begin{array}{cc}
\alpha_k \left\lceil \frac{s_{k}}{\Delta_k} - \frac{1}{2}\right\rfloor +\frac{1}{2} & 1 \le k \le 2K_q\\
\alpha_k\operatorname{l}_{\beta}[s_{k}] & 2K_q < k \le 2K
\end{array}
\right.,
\label{eq:dqlc_mapping}
\end{align}
where $\lceil \cdot \rfloor$ rounds the argument to the nearest integer, $\alpha_k$ is a gain factor which determines the power allocated to each user, $\Delta_k$ is the quantization step for the $k$-th user and $\operatorname{l}_{\beta}[\cdot]$ represents the truncation of the argument to
 $\pm \beta$, with $\beta \in \mathbb{R}$. As observed, the first $K_q$ users transmit a quantized version of their symbols, whereas the $K-K_q$ remaining users simply send a scaled version of their symbols, which will first be truncated if $|s_k|>\beta$. 
 Recall that in the equivalent real-valued model, each user separately encodes the real and imaginary parts of its source symbols. The gain factors must be chosen to ensure that the user power constraints are fulfilled. Assuming Gaussian sources and for a given $\Delta_k$, the power of the quantized symbols for the $k$-th user is given by
\begin{align}
\Gamma(\Delta_k) = 2\sum_{l=0}^{\infty} \left(l + 1/2\right)^2 \left( Q\left( \Delta_k(l+1)\right) - Q\left(\Delta_kl\right) \right),
\label{eq:gamma_function}
\end{align}
where $Q(\cdot)$ is the error function. Hence, the factors $\alpha_k$ should satisfy that $\alpha_k\le \sqrt{\frac{T_k}{\Gamma(\Delta_k)}}, ~1\le k\le 2K_q$. For the uncoded users, $\alpha_k \le \sqrt{T_k}, ~2K_q<k \le 2K$. 

\begin{figure}
	\centering
	\includegraphics[scale=0.9]{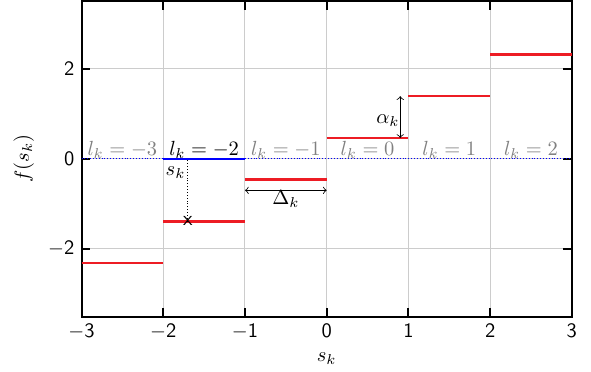}
	\caption{Example of quantized mapping in DQLC with $\Delta_k=1$ and $\alpha_k=0.9$. The source symbol $s_k=-1.7$ is mapped to the interval corresponding to $l_k=-2$.}
	\label{fig:Modulo}
\end{figure}

This mapping function turns out to be suitable for the considered scenarios due to the particular segmentation of the source space carried out by the quantization steps. At the first user, the quantization operation splits its source space into non-overlapping intervals in such a way that all the source values which fall into a particular interval are mapped to the same value. Thus, the transmitted symbol is the interval central point multiplied by the corresponding gain factor $\alpha_1$. According to \eqref{eq:dqlc_mapping}, the distance between the points in the channel space for that user will be $\alpha_1$ (see \Cref{fig:Modulo}). This implies that if the sum of the channel symbols transmitted by the next $K-1$ users and the noise component is lower than $\alpha_1/2$, the received point falls into the same interval and it will be possible to decode the first user correctly. Then, the symbols of the remaining quantized users can be decoded by applying the same idea iteratively. Conversely, when the value resulting from the sum of the user interferences and the noise is larger than $\alpha_1/2$,  the received symbol will cross to an adjacent interval and the decoding procedure breaks down. Therefore,  the key is to properly optimize the parameters $\Delta_k$ and $\alpha_k$ to minimize the distortion between the source and decoded symbols, but lowering the probability of the crossing effect. Note that the use of \ac{DQLC} over a MAC can be actually interpreted as a kind of superposition coding where the information transmitted by the users is weighted by their corresponding channel coefficients and superimposed into the received symbol, whereas the decoding operation could be seen as a kind of successive interference cancellation (SIC). More details about the theoretical aspects of DQLC can be found in \cite{floor15}.

An additional gain can be obtained by choosing an adequate value for the clipping parameter $\beta$ when the correlation between the user symbols is low. In general, the source symbols are unbounded according to the Gaussian pdf. In a practical setup, we would have to design the parameters of the DQLC mapping assuming that the source symbols would fall into a limited interval (for example $[-3,3]$). When this assumption breaks, the decoding fails and the resulting distortion is large for all users. A way to mitigate this effect is to clip the source symbols before transmission.
However, this problem becomes less important as the correlation increases because users transmit more similar symbols, and it is more unlikely that errors are induced on the quantized users. In this paper, we prefer to disregard this step because of a trade-off between the obtained gains and the increase of the complexity in the parameter optimization. On the one hand, the gain provided by the truncation operation is small for medium and high correlation, which is the main focus of the paper. On the other hand, it requires to add a new parameter to the system which should be optimized, which significantly complicates the decoding operation and the design of the optimization algorithm.

As explained in \Cref{sec:model}, the optimal decoding strategy consists in computing the \ac{MMSE} estimates of the source symbols from the received symbol. However, this decoding is unfeasible even for small numbers of users due to the non-linear nature of DQLC. It is hence necessary to resort to lower complexity strategies. Following a similar approach to that presented in \cite{Suarez17} for modulo mappings, we rewrite the non-linear function for the $k$-th quantized user using the following auxiliary function
\begin{align}
f_{l_k}(s_k) = 
\left\{
\begin{array}{cc}
\alpha_k\left(l_k+\frac{1}{2}\right), & s_k\in[\Delta_k l_k, \Delta_k (l_k+1)]\\
\text{0}&\text{otherwise}
\end{array}
\right.,
\label{eq:alt_form_fsk}
\end{align}
where $l_k$ is an integer-valued variable which indexes the quantizer interval where the source symbol $s_k$ falls into. Fig. \ref{fig:Modulo} shows a mapping example for a quantized user with parameters $\Delta_k=1$ and $\alpha_k=0.9$. As observed, each quantizer interval is indexed through its corresponding $l_k$ value such that the above function is only defined for the $l_k$ value corresponding to the interval where the user symbol falls into.
From \eqref{eq:alt_form_fsk}, and also incorporating the uncoded users, we define $\B{f}_{\B{l}}(\B{s}) = [f_{l_1}(s_1),\ldots,f_{l_{2K_q}}(s_{2K_q}),\alpha_{2K_q+1}s_{2K_q+1},\ldots,\alpha_{2K}s_{2K}]^T$, which can also be expressed as
\begin{align}
\B{f}_{\B{l}}(\B{s}) = 
\left\{
\begin{array}{cc}
\B{A}\B{q}_{\B{l}}, & \B{s}\in[\B{a}_{\B{l}}, \B{b}_{\B{l}}]\\
\text{0}&\text{otherwise}
\end{array}
\right.,
\end{align}
with $\B{q}_{\B{l}}=\left[l_1+\frac{1}{2},\ldots,l_{2K_q}+\frac{1}{2},s_{2K_q+1},\ldots, s_{2K}\right]^T$,  $\B{A}=\operatorname{diag}\left\lbrace\alpha_1,\ldots,\alpha_{2K} \right\rbrace$, 
and $\B{l} = [l_1, \ldots, l_{2K_q}]^T$ the vector that stacks the interval indexes for the $K_q$ quantized users. The interval limits are given by $\B{a}_{\B{l}}=[\Delta_1 l_1, \ldots, \Delta_{K_q}l_{K_q}, -\infty, \ldots, -\infty]^T$ and $\B{b}_{\B{l}}=[\Delta_1 (l_1+1), \ldots, \Delta_{K_q}(l_{K_q}+1),\infty,\ldots, \infty]^T$.

 Note that the above equation represents the \ac{DQLC} mapping function for a particular combination of quantization intervals given by the indexes in $\B{l}$. Finally, the mapping function in \eqref{eq:dqlc_mapping} can be expressed as the sum of all functions $\B{f}_{\B{l}}$, i.e.
\begin{align}
\B{f}(\B{s}) = \sum_{\B{l}\in \mathbb{Z}^{2K_q}}
\B{f}_{\B{l}}(\B{s}).
\label{eq:alt_DQLCmapping}
\end{align}

Using \eqref{eq:alt_DQLCmapping}, the computation of the \ac{MMSE} integrals over the whole $2K$-dimensional source space can be restricted to the partition of the source space given by the intervals of the quantized users corresponding to their current source symbols.
However, the number of feasible $\B{l}$ vectors can be arbitrarily large depending on the parameters $\Delta_k$, although it could be delimited according to the source distribution. Since the computational cost of the decoding operation will hence depend on the number of $\B{l}$ vectors to be considered when computing the source estimates, it is essential to design a strategy to lower this number. This can be done with the help of a sphere decoder which selects the most likely $\B{l}$ vectors from the received symbol.

\section{KF-based decoding using the Sphere Decoder}

This section describes the proposed KF-based decoder for \ac{DQLC} mappings which exploits the spatial and the temporal correlation of the source symbols. In a KF setup, the decoding operation consists of a prediction step and an observation step. At the time instant $t-1$, the prediction step generates prior information from the estimates obtained in the corresponding observation step. At the time instant $t$, the observation step computes the estimates of the source symbols using the posterior probability for the received signal and the prior information obtained in the prediction. This iterative procedure is repeated to compute the source estimates at the next instants.  

The a priori information can be computed from \eqref{eq:transition_eq} using the following linear operations
\begin{align}
\bar{\B{s}}_{t|t-1} &= \B{F}\hat{\B{s}}_{t-1}\\
\B{\Sigma}_{t|t-1} &= \B{F}\B{\Sigma}_{t-1}\B{F}^T + \B{C}_w,
\end{align}
where $\bar{\B{s}}_{t|t-1}$ and $\B{\Sigma}_{t|t-1}$ are the mean and the covariance matrix of the a priori information obtained in the prediction step, while $\hat{\B{s}}_{t-1}$ and $\B{\Sigma}_{t-1}$ are the estimate and the error covariance at ~$t-1$. The observation step is however affected by the non-linearities of the \ac{DQLC} function.

For a given prediction $\bar{\B{s}}=\bar{\B{s}}_{t|t-1}$ and the corresponding covariance matrix $\B{\Sigma}_{\B{s}}=\B{\Sigma}_{t|t-1}$, the optimal \ac{MMSE} estimates can be computed as
\begin{align}
\hat{\B{s}}_{\tiny{\text{MMME}}} = \mathbb{E}\left[\B{s}|\B{y}\right] = \frac{\int \B{s} p(\B{y}|\B{s}) p(\B{s}) \text{d}\B{s}}{\int p(\B{y}|\B{s}) p(\B{s}) \text{d}\B{s} },
\label{eq:original_mmse}
\end{align}
where the \ac{pdf} of the source is given by
\begin{align}
p(\B{s}) = \frac{1}{\sqrt{(2\pi)^{2K} \det\{\B{\Sigma}_{\B{s}}\}}} \exp\left(-\frac{1}{2}(\B{s} - \bar{\B{s}})^T \B{\Sigma}_{\mathbf{s}}^{-1} (\B{s} - \bar{\B{s}})
\right),
\label{eq:pdf_source}
\end{align} 
and the conditional probability is
\begin{align}
p(\B{y}|\B{s}) =  \frac{1}{\sqrt{(2\pi\sigma_n^2)}}\exp\left( -\frac{1}{2\sigma_n^2}\Vert \B{y} - \B{H}\B{f}(\B{s})\Vert^2\right).
\label{eq:conditional_prob}
\end{align}
Note that in the real-valued equivalent model, the complex-valued received symbol is represented by a two-element vector $\B{y}$, and the channel vector is transformed into the associated matrix $\B{H}$.

Using the alternative definition of the DQLC mappings in \eqref{eq:alt_DQLCmapping}, the term corresponding to the conditional probability can be expressed as
\begin{align}
p(\B{y}|\B{s}) =\sum_{\B{l}\in\mathbb{Z}^{2K_q}} T\left(\B{y}, \B{H}\B{A}\B{q}_{\B{l}}, \sigma_n^2\B{I}, \B{a}_{\B{l}}, \B{b}_{\B{l}}\right),
\end{align}
where $T(\B{s}, \B{\mu}, \B{C}, \B{a}, \B{b})$ 
represents a truncated Gaussian distribution with mean $\B{\mu}$ and covariance matrix $\B{C}$, in the interval $[\B{a}, \B{b}]$. The vectors $\B{a}_{\B{l}}$ and $\B{b}_{\B{l}}$ are, respectively, the lower and upper limits for the combination of intervals given by $\B{l}$. The above identity allows to rewrite  
\eqref{eq:original_mmse} as
\begin{align}
\scriptsize
\hspace*{-0.1cm}
\hat{\B{s}}_{\tiny{\text{MMME}}} = \frac{\sum_{\B{l}}\int_{a_{\B{l}}}^{b_{\B{l}}} \B{s}  \exp\left( -\frac{1}{2\sigma_n^2}\Vert \B{y} -\B{H}\B{A}\B{q}_{\B{l}}\Vert^2\right) \exp\left(-\frac{1}{2}(\B{s} - \bar{\B{s}})^T \B{\Sigma}_{\B{s}}^{-1} (\B{s} - \bar{\B{s}})\right) \text{d}\B{s}}
{\sum_{\B{l}}\int_{a_{\B{l}}}^{b_{\B{l}}} \exp\left( -\frac{1}{2\sigma_n^2}\Vert \B{y} -\B{H}\B{A}\B{q}_{\B{l}}\Vert^2\right) \exp\left(-\frac{1}{2}(\B{s} - \bar{\B{s}})^T \B{\Sigma}_{\B{s}}^{-1} (\B{s} - \bar{\B{s}})\right) \text{d}\B{s}}.
\label{eq:mmse_porpartes}
\end{align}

On the one hand, the exponent of the conditional probabilities can be rewritten in terms of the variable $\B{s}$ by splitting the contributions corresponding to the quantized users and to the uncoded users, i.e.,
\begin{align}
-\frac1{2\sigma_n^2}\Vert \B{y} -\B{H}\B{A}\B{q}_{\B{l}}\Vert^2  & = - \frac1{2\sigma_n^2}\Vert \B{y} - (\B{H}\B{G}_q\B{q}_{\B{l}} + \B{H}\B{G}_u\B{s})\Vert^2 \\ 
& =  - \frac1{2\sigma_n^2}\Vert \B{y}_l - \B{H}\B{G}_u\B{s} \Vert^2
\label{eq:exponent_conditional}
\end{align}
where $\B{G}_q = \operatorname{diag}\left\lbrace \alpha_1,\ldots, \alpha_{K_q}, 0, \ldots, 0\right\rbrace$,  $\B{G}_u =  \operatorname{diag}\left\lbrace0, \ldots, 0,  \alpha_{K_q+1},\ldots, \alpha_K\right\rbrace$, and $\B{y}_l = \B{y} - \B{H}\B{G}_q\B{q}_{\B{l}}$ represents the remainder after subtracting the symbols transmitted by the quantized users.
Therefore, the two exponents in the integrals of \eqref{eq:mmse_porpartes} can be combined into a single exponent  as a quadratic form in terms of $\B{s}$ as
\begin{align}
\Omega(\B{l}, \B{s}) = \phi_{\B{l}}\exp  \left(-\frac{1}{2} (\B{s} -\B{\mu}_{\B{l}})^T\B{C}_e^{-1}(\B{s}-\B{\mu}_{\B{l}}) \right)
\label{eq:quadratic_form}
\end{align}
where
\begin{align}
\B{\mu}_{\B{l}} &= \B{C}_e \left(\B{\Sigma}_{\B{s}}^{-1}\bar{\B{s}} + \frac1{\sigma_n^2}\B{G}_u^T\B{H}^T\B{y}_l\right)\\ &= 
\bar{\B{s}} + \frac{1}{\sigma_n^2}\B{C}_e \B{G}_u^T\B{H}^T \left(\B{y}_l - \B{H}\B{G}_u\bar{\B{s}} \right)\label{eq:lmmse_for_l}\\
\B{C}_e &= \left( \frac{1}{\sigma_n^2}\B{G}_u^T\B{H}^T\B{H}\B{G}_u + \B{\Sigma}_{\B{s}}^{-1} \right)^{-1}\\
\phi_{\B{l}} &= \exp\left(-\frac1{2}\left(\frac1{\sigma_n^2}\B{y}_l^T\B{y}_l -\B{\mu}_{\B{l}}^T \B{C}_e^{-1}\B{\mu}_{\B{l}}\right)\right). \label{eq:weights}
\end{align}
 The weights $\phi_{\B{l}}$ do not depend on the vector $\B{s}$, and hence the %
 estimates can be computed as
\begin{align}
\hat{\B{s}}_{\tiny{\text{MMME}}} &=\frac{\sum_{\B{l}} \int_{a_{\B{l}}}^{b_{\B{l}}}\B{s}~\Omega(\B{l}, \B{s})~\text{d}\B{s}}{\sum_{\B{l}} \int_{a_{\B{l}}}^{b_{\B{l}}}\Omega(\B{l}, \B{s})~\text{d}\B{s}} \\
&=\frac{\sum_{\B{l}} \phi_{\B{l}}\int_{a_{\B{l}}}^{b_{\B{l}}} \B{s}  \exp  \left(-\frac{1}{2} (\B{s} -\B{\mu}_{\B{l}})^T\B{C}_e^{-1}(\B{s}-\B{\mu}_{\B{l}}) \right) \text{d}\B{s}}
{\sum_{\B{l}} \phi_{\B{l}} \int_{a_l}^{b_l}\exp  \left(-\frac{1}{2} (\B{s} -\B{\mu}_{\B{l}})^T\B{C}_e^{-1}(\B{s}-\B{\mu}_{\B{l}}) \right) \text{d}\B{s}},
\label{eq:alt_mmse}
\end{align}
 where the original integrals in \eqref{eq:original_mmse} are replaced by a set of integrals over the resulting truncated Gaussian functions in their corresponding intervals. The above expression also provides the optimal \ac{MMSE} estimates for the considered scenario, but it requires to numerically compute as many integrals as the number of $\B{l}$ vectors we consider. It is hence important to find a set of feasible vectors as small as possible in order to make the decoding operation affordable. The original problem of estimating the source symbols is hence transformed into the search of the most likely discrete vectors $\B{l}$ used during the transmission.

Unlike in the case of modulo functions \cite{Suarez17}, large values of the weights $\phi_{\B{l}}$ do not necessarily imply that the vector $\B{l}$ corresponds to a likely set of quantizer intervals for the current source vector. For this reason, the search of the most likely $\B{l}$ vectors must be carried out over the whole expression in \eqref{eq:quadratic_form} which depends on $\B{l}$ and $\B{s}$. An ideal approach to determine the relevant $\B{l}$ vectors would be to have a closed-form expression for the maximum of $\Omega(\B{l}, \B{s})$ only as a function of $\B{l}$. Although this is not possible, we can circumvent this limitation by evaluating the truncated Gaussian functions in a representative point which is expected to provide large values for likely vectors $\B{l}$.
In particular, for a given vector $\B{l}$, the evaluation point is chosen as the vector consisting of the middle point of the corresponding intervals for the quantized users and the linear MMSE estimates for the uncoded users. Thus, we define the $2K$-dimensional point 
\begin{align}
\tilde{\B{s}}_{\B{l}} = \left[\B{D}\left(\B{l} + \frac{1}{2}\B{1}\right), \left[\B{\mu}_{\B{l}}\right]_{2K_q+1:2K} \right]^T,
\label{eq:s_tilde}
\end{align}
 with $\B{D}=\operatorname{diag}\left\lbrace\Delta_1,\ldots,\Delta_{2K_q} \right\rbrace,$
and the problem changes into finding the $\B{l}$ vectors that satisfy
\begin{align}
\Omega(\B{l}, \tilde{\B{s}}_{\B{l}}) > R',
\label{eq:lattice_condition}
\end{align}
where $R'$ is a given threshold. Hence, we define the set of feasible $\B{l}$ vectors as 
\begin{align}
\mathcal{L}_R = \{\B{l}\in \mathbb{Z}^{2K} ~|~~ \Omega(\B{l}, \tilde{\B{s}}_{\B{l}}) > R'\}.
\end{align}

Since the last components of $\tilde{\B{s}}_{\B{l}}$ are equal to the linear MMSE estimates in \eqref{eq:lmmse_for_l}, the term $\tilde{\B{s}}_{\B{l}}-\B{\mu}_{\B{l}}$ is zero in the last $2K-2K_{q}$ components corresponding to the uncoded users. This allows to express the exponent of each truncated Gaussian only as a function of $\B{l}$. %
Following an approach similar to \cite{Suarez17}, the expression for $\Omega(\B{l}, \tilde{\B{s}}_{\B{l}})$ can be rewritten in a lattice form, thus enabling the use of a sphere decoder to search the set of $\B{l}$ vectors that satisfies the condition in \eqref{eq:lattice_condition}. 

\begin{lemma}
	The term $\Omega(\B{l}, \tilde{\B{s}}_{\B{l}})$ can be rewritten in a lattice form as 
	\begin{align}
	\Omega(\B{l}) = \exp\left(-\frac1{2}\left(\B{l} - \B{l}_o \right)^T\B{\Lambda}\left( \B{l} - \B{l}_o \right)\right), 
	\label{eq:final_lattice}
	\end{align}
\end{lemma}
where
\begin{align}
\B{\Lambda} &= \B{B}^T \tilde{\B{C}}_e^{-1}\B{B} + \B{A}_q^T\B{H}_q^T\B{Z}\B{H}_q\B{A}_q
\label{eq:lattice_matrix}
\end{align}
is the matrix that represents the lattice that models the search space and~
$\B{l}_o = \Lambda^{-1}\left(\B{B}^T\tilde{\B{C}}_e^{-1}\B{v} + \B{m}\right)$
represents the centre of the sphere where candidates for $\B{l}$ will be searched. The auxiliary vectors and matrices required in the expressions above are given by
\begin{align}
\B{B} &= \B{D} + \frac{1}{\sigma_n^2}\B{C}_c \B{A}_u^T\B{H}_u^T\B{H}_q\B{A}_q \label{eq:matrixB},\\
\B{Z} &= \left(\sigma_n^2\B{I} + \B{H}\B{G}_u\B{\Sigma}_{\B{s}}\B{G}_u^T\B{H}^T\right)^{-1},\\
\B{v} &= \frac{1}{\sigma_n^2}\B{C}_c \B{A}_u^T\B{H}_u^T \left(\B{y} - \B{H}_q\B{A}_q\frac1{2}\B{1} -\B{H}_u\B{A}_u\bar{\B{s}}^{(q)}\right) -
\frac{1}{2}\B{D}\B{1} +\bar{\B{s}}^{(q)},\\
\B{m} &= \B{A}_q^T\B{H}_q^T\left(\frac{1}{\sigma_n^2}\B{H}\B{G}_u\B{C}_e\B{\Sigma}_{\B{s}}^{-1}\bar{\B{s}} + \B{Z}\left(\B{y} -\frac1{2}\B{H}_q\B{A}_q\B{1}\right)\right),
\end{align}
with $\B{H} = [\B{H}_q, \B{H}_u]$, $\B{A}_q = \operatorname{diag}\left\{ \alpha_1,\ldots,\alpha_{K_q} \right\}$ and $\B{A}_u = \operatorname{diag}\left\{ \alpha_{K_{q}+1},\ldots,\alpha_{K} \right\}$. All the steps required to obtain the expression above can be found in the appendix.

Using this lattice, the search of the relevant vectors $\B{l}$ satisfying \eqref{eq:lattice_condition} is equivalent to finding those vectors whose corresponding exponent is below a given threshold, i.e,
\begin{align}
\frac1{2}\left(\left(\B{l} - \B{l}_o \right)^T\B{\Lambda}\left( \B{l} - \B{l}_o \right)\right) < R,
\label{eq:condition}
\end{align}
where $R$ represents the radius of the sphere that contains the relevant $\B{l}$ vectors.  Now, a sphere decoder can be used to efficiently determine the $\B{l}$ points in the lattice that fall into a sphere of radius $R$ centered in $\B{l}_o$ \cite{Hochwald03}. The application of the sphere decoder to carry out this search is similar to that described in \cite{Suarez17}.  Starting with the Cholesky decomposition of the lattice, $\B{\Lambda} = \B{L}\B{L}^T$, where $\B{L}$ is a lower triangular matrix, we can pose an iterative algorithm that, at each iteration, selects the range of integer values for each component of $\B{l}$ that satisfies the considered radius taking into account the previous components. After $2K$ iterations, the algorithm will only provide the $\B{l}$ vectors whose corresponding exponent in \eqref{eq:final_lattice} is lower than $R$. %

\begin{algorithm}
	\begin{algorithmic}
		\STATE $\bar{\B{s}}_{1|0}\gets\B{0},\B{\Sigma}_{1|0}\gets\B{C}_s$
		\STATE $\B{A}_1\gets \operatorname{diag}(\alpha_1,\ldots,\alpha_{2K}),\B{D}_1\gets\operatorname{diag}(\Delta_1,\ldots,\Delta_{2K_q})$ for $\B{\Sigma}_{1|0}$
		\FORALL{$t \in [1, T]$}
		\STATE $\B{C}_e \gets \left( \frac{1}{\sigma_n^2}\B{G}_u^T\B{H}^T\B{H}\B{G}_u + \B{\Sigma}_{t|t-1}^{-1} \right)^{-1}$
		\STATE $\B{\mu}_{\B{l}}\gets\bar{\B{s}}_{t|t-1} + \frac{1}{\sigma_n^2}\B{C}_e \B{G}_u^T\B{H}^T \left(\B{y}_l - \B{H}\B{G}_u\bar{\B{s}}_{t|t-1} \right)$
		\STATE Build the set $\mathcal{L}_R$ using the sphere decoder
		\FORALL{$\B{l} \in \mathcal{L}_R$}
		\STATE $ \phi_{\B{l}} \gets \exp\left(-\frac1{2}\left(\frac1{\sigma_n^2}\B{y}_l^T\B{y}_l -\B{\mu}_{\B{l}}^T \B{C}_e^{-1}\B{\mu}_{\B{l}}\right)\right)$
		\STATE $\B{m}_{\B{l}} \gets \mathbb{E}[\B{s}], ~~ \B{s} \sim T(\B{s},\B{\mu}_{\B{l}}, \B{C}_e,a_{\B{l}}, b_{\B{l}})$%
		\STATE $\tau_{\B{l}} \gets \int T(\B{s},\B{\mu}_{\B{l}}, \B{C}_e,a_{\B{l}}, b_{\B{l}}) ~d\B{s}$
		\STATE $\B{\Sigma}_{\B{l}}\gets \operatorname{Cov}[\B{s}], ~\B{s} \sim T(\B{s},\B{\mu}_{\B{l}}, \B{C}_e,a_{\B{l}}, b_{\B{l}})$
		\ENDFOR
		\STATE $\hat{\B{s}}_{t} \gets \sum_{\B{l}\in\mathcal{L}_R} \phi_{\B{l}} \B{m}_{\B{l}}~/ \sum_{\B{l}\in\mathcal{L}_R} \phi_{\B{l}} \tau_{\B{l}}$
		\STATE $\B{\Sigma}_t \gets \sum_{\B{l}\in\mathcal{L}_R}  \phi_{\B{l}} \B{\Sigma}_{\B{l}}~/\sum_{\B{l}\in\mathcal{L}_R} \phi_{\B{l}} \tau_{\B{l}}$
		\STATE $\bar{\B{s}}_{t+1|t} \gets \B{F}\hat{\B{s}}_{t}$
		\STATE $\B{\Sigma}_{t+1|t} \gets \B{F}\B{\Sigma}_{t}\B{F}^T + \B{C}_w$
		\STATE Update $\B{A}_{t+1}, \B{D}_{t+1}$ solving \eqref{final_problem} with $\B{\Sigma}_{t+1|t}$\\
		\ENDFOR
	\end{algorithmic}
	\caption{Algorithm to compute the source estimates from the \ac{KF}-based decoder.}
	\label{alg:kf}
\end{algorithm}

Next, a MMSE estimate of the source symbols is obtained by applying \eqref{eq:alt_mmse} with the set of vectors provided by the sphere decoder. Finally, the covariance matrix $\B{\Sigma}$ is updated using the estimates obtained in the observation step, and the DQLC parameters are optimized for the ensuing time instant as described in \Cref{sec:parameters}. \Cref{alg:kf} summarizes the set of steps performed by the proposed algorithm to compute the estimates of the source symbols at each time instant using the proposed KF-based decoder and the sphere decoder.

Note that if the radius $R$ of the sphere decoder is adequately chosen, the number of candidate vectors will be small and the computational complexity of the decoding operation will be reduced significantly with respect to solving the original integrals in \eqref{eq:original_mmse}. In a similar way, the value of the radius should be selected to ensure that those $\B{l}$ vectors with a significant weight in the computation of the \ac{MMSE} estimates fall into the corresponding sphere. In the next section, we explain how to design the radius of the sphere decoder and its impact on the computational cost. %

\subsection{Radius of the Sphere Decoder}\label{sec:radius_sd}

In this section, we estimate the minimum radius required to detect the optimal vector $\B{l}^*$, i.e., the vector that includes the actual combination of quantized intervals used in transmission. %
Assuming we know the vector $\B{l}^*$,  we can determine the distribution of the corresponding points in the lattice given by \eqref{eq:final_lattice} as a function of the potential source vectors and the noise. %
 Considering the exponents of $\phi_{\B{l}}$ in \eqref{eq:weights}, and replacing $\B{y}_l$ and $\B{\mu}_{\B{l}}$ by the received signal for  $\B{l}^*$, we obtain
\begin{align*}
 \frac1{\sigma_n^2}&\B{y}_{l^*}^T\B{y}_{l^*} -\B{\mu}_{\B{l}^*}^T \B{C}_e^{-1}\B{\mu}_{\B{l}^*} = \frac{1}{\sigma_n^2}(\B{H}\B{G}_u\B{s}+\B{n})^T(\B{H}\B{G}_u\B{s}+\B{n}) ~- \notag \\
&\left(\bar{\B{s}} + \frac{1}{\sigma_n^2}\B{C}_e\B{G}_u^T\B{H}^T(\B{H}\B{G}_u(\B{s}-\bar{\B{s}}) + \B{n})\right)^T\B{C}_e^{-1}  ~~\times \notag \\  
& \hspace*{3.2cm} \left(\bar{\B{s}} + \frac{1}{\sigma_n^2}\B{C}_e  \B{G}_u^T\B{H}^T(\B{H}\B{G}_u(\B{s}-\bar{\B{s}}) + \B{n})\right)\notag\\
\end{align*}
\vspace*{-0.8cm}
\begin{align*}
= ~ &\frac{1}{\sigma_n^2}(\B{H}\B{G}_u(\B{s}-\bar{\B{s}})+\B{n}+\B{H}\B{G}_u\bar{\B{s}})^T(\B{H}\B{G}_u(\B{s}-\bar{\B{s}})+\B{n}+\B{H}\B{G}_u\bar{\B{s}})~ - \notag \\
& \left(\B{C}_e^{-1}\bar{\B{s}} + \frac{1}{\sigma_n^2}\B{G}_u^T\B{H}^T(\B{H}\B{G}_u(\B{s}-\bar{\B{s}}) + \B{n})\right)^T\B{C}_e ~~\times \notag\\
& \hspace*{2.9cm}\left(\B{C}_e^{-1}\bar{\B{s}} + \frac{1}{\sigma_n^2}\B{G}_u^T\B{H}^T(\B{H}\B{G}_u(\B{s}-\bar{\B{s}}) + \B{n})\right)\notag\\
\end{align*}
\vspace*{-0.8cm}
\begin{align*}
= ~
&(\B{H}\B{G}_u(\B{s}-\bar{\B{s}})+\B{n})^T~\B{Z}~(\B{H}\B{G}_u(\B{s}-\bar{\B{s}})+\B{n}) - \bar{\B{s}}^T\B{\Sigma}_s^{-1}\bar{\B{s}},
\end{align*}

Given that $\B{G}_u$ is a matrix with zeros except on its last $2K_q$ diagonal elements, this exponent follows a chi-squared distribution with $2K_q$ degrees of freedom.

On the other hand, the other exponent in the lattice comes from the expression $(\B{s}_{\B{l}}-\B{\mu}_{\B{l}})^T\B{C}_e^{-1}(\B{s}_{\B{l}}-\B{\mu}_{\B{l}})$ which is chi-squared distributed with $2K$ degrees of freedom, and it can be rewritten as
\begin{align*}
&(\B{s}-\B{\mu}_{\B{l}})^T\B{C}_e^{-1}(\B{s}-\B{\mu}_{\B{l}}) 
= \notag\\
&\left(\B{s}- \bar{\B{s}} - \frac{1}{\sigma_n^2}\B{C}_e\B{G}_u^T\B{H}^T(\B{H}\B{G}_u(\B{s}-\bar{\B{s}}) - \B{n}) \right)^T\B{C}_e^{-1} ~~ \times \notag \\ 
&\hspace*{2.7cm}\left(\B{s} - \bar{\B{s}} - \frac{1}{\sigma_n^2}\B{C}_e\B{G}_u^T\B{H}^T(\B{H}\B{G}_u(\B{s}-\bar{\B{s}}) - \B{n}) \right)\notag\\
\end{align*}
\vspace*{-0.8cm}
\begin{align*}
&=  \left( \B{\Sigma}_s^{-1}(\B{s}-\bar{\B{s}}) - \frac{1}{\sigma_n^2}\B{G}_u^T\B{H}^T\B{n} \right)^T\B{C}_e \left( \B{\Sigma}_s^{-1}(\B{s}-\bar{\B{s}}) - \frac{1}{\sigma_n^2}\B{G}_u^T\B{H}^T\B{n} \right).\notag
\end{align*}

\vspace*{0.2cm}
Also, the Gaussian vectors are uncorrelated with each other
\begin{align}
&E\left[(\B{H}\B{G}_u(\B{s}-\bar{\B{s}})+\B{n})\left( \B{\Sigma}_s^{-1}(\B{s}-\bar{\B{s}}) - \frac{1}{\sigma_n^2}\B{G}_u^T\B{H}^T\B{n} \right)^T\right] =  \notag\\
& = \B{H}\B{G}_u - \B{H}\B{G}_u = \B{0}.
\end{align}

Hence, the sum of the exponents corresponds to a chi-squared distribution of $2K+2K_q$ degrees of freedom. This implies that, assuming the index vector $\B{l}^*$ was employed at transmission, the values for the exponent of $\Omega(\B{l}^*)$ in \eqref{eq:final_lattice} will follow a chi-squared distribution with $2K+2K_q$ degrees of freedom, i.e,
\begin{equation}
\left(\B{l}^* - \B{l}_o \right)^T\B{\Lambda}\left( \B{l}^* - \B{l}_o \right) \sim \chi^2_{2K+2K_q}.
\end{equation}

According to \eqref{eq:condition}, we must choose the radius $R$ to guarantee that the potential values for the above exponent are lower than $2R$ with a high probability, i.e,
\begin{equation}
g_{2K+2K_q}(2R) \geq  1-\tau,
\end{equation}
where $g_n(x) = P(\chi_{n}^2 < x)$ is the cumulative density function of a chi-squared variable with $n$ degrees of freedom and $\tau$ represents an arbitrarily small probability. Hence, a useful criterion to select the sphere radius is given by
\begin{align}
R \ge g^{-1}_{2K+2K_q}(1-\tau)/2,
\label{eq:minimum_radius}
\end{align}
where $g^{-1}_k (\cdot)$ is the inverse function of the  cumulative density function for a chi-squared distribution. The term, $1-\tau$ determines the probability of the vector $\B{l}^*$ to fall into a sphere of radius $R$ and centre $\B{l}_o$. It is hence important to reach a trade-off for this parameter to avoid failing at the decoding operation without increasing unnecessarily the radius.
We have checked experimentally that $\tau  \approx 10^{-4}$ provides a good behaviour, although if no candidates are found for a given $\tau$, decoding could be repeated with a larger parameter until solutions are found.

\subsection{Analysis of the Computational Cost}

The computational cost of the proposed \ac{DQLC}-based scheme mainly depends on the decoding operation because the cost of applying the encoding function is negligible. 
Using the proposed decoder in \Cref{alg:kf}, the computation of the \ac{MMSE} estimates at the receiver involves to solve as many integrals of truncated Gaussian distributions as the number of vectors $\B{l}$ in the set $\mathcal{L}_R$ built with the sphere decoder. Solving the integrals corresponding to a truncated Gaussian has some advantages respect to computing the original \ac{MMSE} integrals in \eqref{eq:original_mmse}. %
Firstly, efficient techniques can be applied to numerically solve these integrals \cite{Genz09}. Also, the number of samples required to apply Monte Carlo techniques is much lower than in the case of the original integrals since the size of the integration intervals is delimited by the quantization steps $\Delta_k$ which are in general small. However, these computational benefits vanish as the size of the set $\mathcal{L}_R$ increases.  

A general analysis of the sphere decoder algorithm is provided in \cite{Hassibi05}, and it relies on estimating the number of lattice points enclosed by a sphere of a given radius. In our case, the number of vectors which fall into the sphere will be larger as the number of users increases and when the source correlation tends to zero since, in those cases, the uncertainty in the decoding operation is also larger. However, if the radius $R$ %
is properly chosen and the encoding parameters are optimized to minimize the ambiguities in the decoding operation (see \Cref{sec:parameters}), the size of $\mathcal{L}_R$ should remain small enough to guarantee an acceptable overall complexity.  %

\section{Parameter Optimization} \label{sec:parameters}

An important issue to improve the performance of the proposed DQLC-based scheme is the optimization of the mapping parameters $\alpha_k$ and $\Delta_k$ for each \ac{MAC} user. Since an exhaustive search over the parameter space becomes prohibitive as the number of users increases, we propose to optimize $\alpha_k$ and $\Delta_k$ according to the following constrained optimization problem
\begin{align}
\underset{\B{D},\B{A}}{\arg\min} &~~ E\left[ \left|\B{s} - \hat{\B{s}}_{\text{MMSE}}\right|^2 \right]
\label{eq:original_mse_minimization}\\
s.t.&~~ 0 \le \alpha_k \le \sqrt{\frac{T_k}{\Gamma(\Delta_k)}}, ~\qquad\forall k,1 \le k \le 2K_q \notag \\
&~~ 0 \le \alpha_k \le \sqrt{T_k}, ~~~~~~\qquad\forall k, 2K_q< k \le 2K \notag,
\end{align}
where $\B{D}$ and $\B{A}$ are diagonal matrices that contain the parameters $\Delta_k$ and $\alpha_k$ for the $K$ users. As commented in \Cref{sec:dqlc}, the key point in the design of the DQLC parameters is to guarantee the quantized symbols do not go across the adjacent intervals due to the information transmitted by near users. %
Thus, the DQLC parameters must be chosen to minimize the average distortion while ensuring that the quantized users are correctly decoded. This can be accomplished by adding two additional constraints to the above problem.
First, a necessary condition that the quantized users must satisfy is that the diagonal elements of the matrix resulting from the Cholesky decomposition of $\B{\Lambda}$ is larger than a certain value. As shown in \cite{Suarez17}, this constraint ensures that the points on the lattice space are sufficiently far away from one another which, in practice, implies that the probability of crossing to a wrong interval will be lower.

Another important issue is the fact that allocating more power to a user in DQLC implies, in general, to increase the $\Delta_k$ parameters for previous users and, consequently, the $\alpha_k$ parameters to avoid ambiguities in the received channel symbols. However, the maximum value for $\alpha_k$ is upper bounded by the available power. Hence, if this bound is reached by some users, allocating more power to others will cause ambiguities in the decoding process.
In order to avoid this situation, we introduce a constraint over the maximum achievable values for the $\alpha_k$ of the quantized users. 
From \eqref{eq:gamma_function}, it is straightforward to see that  
\begin{equation}
\lim_{\Delta_k \rightarrow\infty} \Gamma(\Delta_k) \approx \frac{1}{2}
\end{equation}
and hence, given that $\alpha_k\le \sqrt{\frac{T_k}{\Gamma(\Delta_k)}}$, it is required to ensure that the $\alpha_k$ values corresponding to the quantized users remain below $\sqrt{2T_k}$, $ \forall k=1,\ldots,2K_q$. If this threshold were reached for some quantized user, that would imply the need of allocating more power to the previous users to avoid ambiguities in the decoding operation, but this is not possible without violating their power constraint.

Considering these two constraints, the initial optimization problem in \eqref{eq:original_mse_minimization} is approximated as
\begin{align}
\underset{\B{D},\B{A}}{\arg\min} &~~e(\B{D},\B{A})
\label{eq:original_mse_minimization_simp}&\\
s.t.&~~ 0 \le \alpha_k \le \sqrt{\frac{T_k}{\Gamma(\Delta_k)}}, &\forall k,1 \le k \le 2K_q \notag \\
&~~ 0 \le \alpha_k \le \sqrt{T_k}, &\forall k, 2K_q< k \le 2K \notag\\
&~~ 0 \le \alpha_k \le \sqrt{2T_k}-\mu, &\forall k,1 \le k \le 2K_q \label{eq:far_constraint}\\
&~~ [\B{L}]_{k,k}\ge S\label{eq:diagonal_constraint},&
\end{align}
where $e(\B{D},\B{A})$ is the error assuming that the intervals of the quantized symbols are correctly guessed at the receiver,  $\mu$ avoids that the $\alpha_k$ values of the quantized users achieve their maximum value for a  large $\Delta_k$, and  $[\B{L}]_{k,k}$ represents the $k$-th diagonal element of the matrix resulting from the Cholesky decomposition of the lattice. The parameter $S$ is a constant to ensure those diagonal elements are above some threshold and it is of the same order of magnitude as the radius $R$.

The cost function of the above problem, $e(\B{D},\B{A})$, consists of two different contributions to the overall distortion%
: the quantization errors and the distortion observed in the uncoded symbols. We first obtain an upper bound for the error of the quantized users as 
\begin{align}
e_{q,k}(\Delta_k) = & \sum_{i=1}^{\infty}\int_{\Delta_ki}^{\Delta_k(i+1)} (s - \delta_i)^2p(s)ds,
\end{align}
where $\delta_i$ is the decoded value for the user transmitted in the $i$-th interval and is given by
\begin{align}
\delta_i =& \int_{a_i}^{b_i} s p(s)ds =  \sqrt{\frac{2\sigma_s^2}{\pi}}\frac{\exp(-a_i^2) - \exp(-b_i^2)}{Q(b_i) - Q(a_i)}
\end{align}
with $a_i = \Delta_k i$ and $b_i = \Delta_k (i+1)$. Hence, the above bound can be expressed as
\begin{align}
e_{q,k}(\Delta_k) = & 1 + \frac{1}{2}\sum_{i=1}^{\infty} \delta_i^2(Q(b_i) - Q(a_i)) \notag\\
-& \sqrt{\frac{2\sigma_s^2}{\pi}}\sum_{i=1}^{\infty} \delta_i(\exp(-a_i^2) - \exp(-b_i^2)).
\end{align}
Note that this is an upper bound since it does not consider the source correlation. 
Then, an upper bound for the error of the uncoded users is computed as $e_{u,k}(\B{A}_u) = [\B{C}_e]_{k,k}$. Finally, an upper bound on the overall MMSE assuming the quantized users are correctly decoded is given by
\begin{align}
e(\B{D}, \B{A}_u) = \sum_{k=1}^{2K_q} e_{q,k}(\Delta_k) + \sum_{k=2K_q+1}^{2K}e_{u,k}.
\end{align}

We now address the rewriting of the constraint \eqref{eq:diagonal_constraint}. %
At the $k$-th user and for a given $\Delta_k$, the parameter $\alpha_k$ is given by $\alpha_k=\frac{p_k}{\sqrt{\Gamma(\Delta_k)}}$, where $p_k$ is the power allocated to that user and it must satisfy $|p_k|^2 \le T_k$. In addition, $\Delta_k \approx \sqrt{\frac{1}{\Gamma(\Delta_k)}}$ for low $\Delta_k$ values and hence the parameter $\alpha_k$ can be approximated as $\alpha_k\approx p_k\Delta_k$. Considering this alternative definition of the $\alpha_k$ parameters, 
we decompose the diagonal matrix $\B{A}_q$ as $\B{A}_q = \B{P}_q\B{D}$,  with $\B{P}_q=\operatorname{diag}\left\{ p_1, \ldots, p_{2K_q} \right\}$.  
Replacing this approximation in the lattice expression given by \eqref{eq:lattice_matrix}, we obtain
\begin{align}
\B{\Lambda} \approx \B{D}\left(\bar{\B{B}}^T \tilde{\B{C}}_e^{-1}\bar{\B{B}} + \B{P}_q^T\B{H}_q^T\B{Z}\B{H}_q\B{P}_q
\right)\B{D}
\end{align}
with
\begin{align}
\bar{\B{B}} = \left(\mathbf{I}+ \frac{1}{\sigma_n^2}\B{C}_c \B{A}_u^T\B{h}_u\B{h}_q^T\B{P}_q\right). \notag
\end{align}
We now define the lattice $$\bar{\B{\Lambda}} = \B{D}^{-1}\B{\Lambda}\B{D}^{-1}=\B{D}^{-1}\B{L}^T\B{L}\B{D}^{-1},$$ that only depends on $\B{P}_q$ and $\B{A}_u$. On the other hand, the parameter $\alpha_k$ is simply a factor scale for uncoded users, and therefore we can define $\B{P}_u = \B{A}_u = \operatorname{diag}\{\alpha_{2K_q+1}, \ldots, \alpha_{2K}\}$.  
Hence, for given power allocations $\B{P}_q$ and $\B{P}_u$, we can determine the minimum $\Delta_k$ that ensures the diagonal elements of $\B{L}$ are above some threshold $S$. This can be computed with the help of the decomposition $\bar{\B{\Lambda}} = \bar{\B{L}}^T\bar{\B{L}}$ and making
\begin{align}
\Delta_k = \frac{S}{\left[\bar{\B{L}}\right]_{k,k}}, ~~\forall k, ~1 \le k \le 2K_q.
\end{align}
Replacing the constraint \eqref{eq:diagonal_constraint} by this expression, and taking into account that  $\alpha_k=p_k\sqrt{\frac{1}{\Gamma(\Delta_k)}}, \forall k, 1\le k \le 2K_q$, the problem \eqref{eq:original_mse_minimization} is transformed into
\begin{align}
\underset{\B{P}_q,	\B{P}_u}{\arg\min} ~~& \sum_{k=1}^{K_q} e_{q}(\Delta_k) + \sum_{k=K_q+1}^{2K}e_{u,k} & \label{final_problem}\\
s.t.~~& 0 \le p_k \le \sqrt{T_k}, & \forall k\notag\\
~~&\frac{p_k}{\sqrt{\Gamma(\Delta_k)}} \le \sqrt{2T_k}-\mu, &\forall k, 1\le k \le 2K_q,\notag\\
~~& \Delta_k = \frac{S}{[\bar{\B{L}}]_{k,k}}, &\forall k, 1\le k \le 2K_q, \notag
\end{align}
which searches the optimal power allocations for the $K$ users considering that the $\Delta_k$ values are directly determined from such power allocations. This is a non-linear optimization problem that must be solved numerically, but the computation of the cost function and the constraints have a lower complexity than the exact computation of the expected distortion \cite{floor15}. %
Finally, the search space is reduced since the quantization steps $\Delta_k$ are estimated from the user power allocations. %

As shown in \Cref{alg:kf}, the covariance matrix $\B{\Sigma}$ is updated after the observation step by using the obtained estimates and the a priori information. The mapping parameters are then optimized at the receiver by using the new covariance matrix and the resulting values are fed back to the users which will encode the next source symbol with the optimized DQLC scheme.

\section{Results}

In this section, the results of several computer experiments are presented to illustrate the performance of the proposed \ac{DQLC}-based scheme for different fading MAC scenarios. At each time instant, a vector of $K$ source symbols is generated from the autoregressive model described by \eqref{eq:transition_eq}. In particular, we assume a correlation model where $\B{F} = \varphi\B{I}$ and $\B{C}_w=\left(1-\rho^2\right)\B{C}_s$, with $0 < \varphi < 1$ and $0 < \rho < 1$ scalar terms that determine the level of temporal and spatial correlation, respectively. According to this model, the vectors of source symbols follow a multivariate circularly symmetric complex-valued Gaussian distribution with zero mean and covariance matrix $\B{C}_{\B{s}}$. Unless explicitly mentioned, we focus on a spatial correlation model where $[\B{C}_s]_{i,i} = 1$ and $[\B{C}_s]_{i,j} = \rho, ~\forall i,j, i\neq j$.

The $K$ source symbols are then encoded using \ac{DQLC} with the parameters provided by the receiver through the feedback channel. %
After the encoding operation, the resulting symbols are sent over a block fading MAC. The channel response is assumed to remain static during the transmission of a block of $T$ consecutive symbol vectors, but it varies from one block to the other. The different channel realizations are assumed to follow a Rayleigh distribution. Without loss of generality, we assume that the channels are real, because if a channel has an imaginary part, the optimal precoding strategy consists in multiplying the user symbols by a complex-valued gain which cancels out the channel phase \cite{Suarez18b}.  Also, we consider $|h_1|>|h_2|>\ldots>|h_K|$ since, as commented, the user channel responses are assumed to be known at the receiver and it could use that information to reorder the channel gains appropriately. The received signal is employed to compute an estimate of the current source symbols with the proposed KF-based decoder and with the help of the sphere decoder. Finally, the receiver updates the a priori information, determines the optimal values for the \ac{DQLC} parameters and feeds back this information to the users. 

At each computer experiment,  we consider the transmission of blocks of $T=100$ source vectors over $L=2000$ different channel realizations. Results are averaged over all channel realizations. The value $T=100$ is chosen to show that the decoder does not diverge after the successive decoding of the received samples. However, blocks of a smaller size $(10<T<100)$ provide similar results due to the fast convergence of the decoding procedure. %
The performance of the transmission scheme is measured according to the average \ac{MSE} between the source and estimated symbols, which in this case is empirically calculated as
 \begin{align}
 \xi = \frac1{LTK}\sum_{l=1}^L\sum_{t=1}^T\sum_{k=1}^K |s_{t,k}^{(l)} - \hat{s}_{t,k}^{(l)}|^2,
 \end{align}
 where $s_{t,k}^{(l)}$ represents the source symbol of the $k$-th user at the time instant $t$ in the $l$-th block, and $\hat{s}_{t,k}^{(l)}$, its corresponding estimate.
  In this section, the figures with the obtained results will show the \ac{SDR} obtained for a given range of \acp{SNR}, where the \ac{SDR} is defined as
 $\text{SDR} = 10\log_{10}\left(1/\xi\right)$,
and the SNR for the $k$-th user is defined as $\eta_k = 10\log_{10}(T_k)$. Thus, we assume that noise variance is equal to $1$ and the SNRs are given directly by the power constraints. For simplicity, we focus on a scenario where $\eta_k = \eta ~\forall k$. 

\begin{figure}
	\centering
	\includegraphics[scale=0.9]{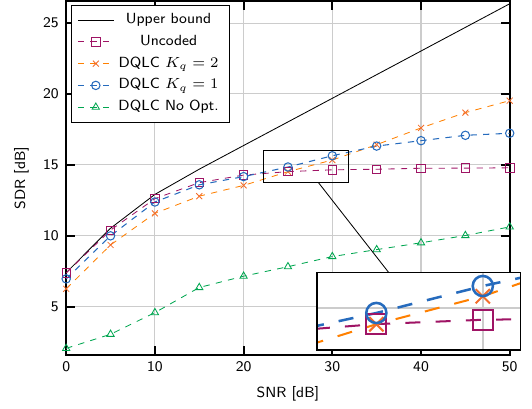}
	\caption{Performance of the DQLC-based and uncoded schemes for $K=3$ users, with $\varphi=0$ and a spatial correlation $\rho=0.95$.}
	\label{fig:FigSdr_3-1x1a}
\end{figure}

\begin{figure}
	\centering
	\includegraphics[scale=0.9]{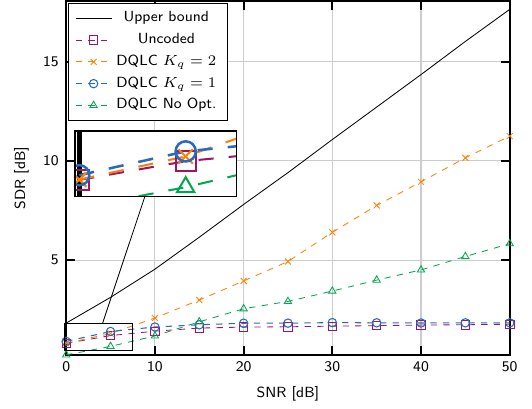}
	\caption{Performance of the different DQLC-based systems and the uncoded scheme for $K=3$ users and considering uncorrelated symbols in both the spatial domain and the temporal domain, i.e. $\varphi=\rho =0$.}
	\label{fig:FigSdr_3-1x1b}
\end{figure}

In the first experiment, we consider a \ac{MAC} scenario with $K=3$ users whose source symbols are uncorrelated in the time domain, i.e., $\varphi = 0$. Thus, the decoder does not have a priori information to improve the estimation of the source symbols. \Cref{fig:FigSdr_3-1x1a} shows the performance of the \ac{DQLC}-based scheme for a correlation factor $\rho=0.95$ and considering two different configurations depending on the number of quantized users: $K_q=2$ and $K_q=1$. We also include the performance of a \ac{DQLC} scheme which use the same set of parameters regardless of the channel realization or the \ac{SNR} value. In particular, we choose $\Delta_1=\Delta_2=1$ and $\B{\alpha} = [1; 0.2; 0.025]$ to ensure a correct decoding avoiding to break down the DQLC system.  
The performance of the different \ac{DQLC} systems are compared to that of a linear scheme where the users send a scaled version of their data. In this case, the complex-valued scale factors are adjusted to allocate the optimal power to each user \cite{Suarez18b}. Finally, a performance upper bound is also included in both figures as benchmark. This bound is computed following a similar argument to \cite[Proposition IV.1]{lapidoth10}, such that 
the rate distortion function for multivariate Gaussian sources is equated to the sum-capacity of the channel under the assumption of user collaboration and a power boost provided by the source correlation in the MAC.
 \Cref{fig:FigSdr_3-1x1b} shows the same results for a correlation factor $\rho=0$ (uncorrelated symbols).

As observed, the \ac{DQLC} system with $K_q=2$ outperforms the linear scheme from a given \ac{SNR} value that depends on the spatial correlation. For a correlation factor $\rho=0.95$, linear and DQLC systems provide similar SDR values for $\eta \le 25$ dB. However, the gain of DQLC is more perceptible for high SNRs where it becomes about $5$ dB for $\eta=50$ dB. For low correlation factors, the performance of the linear system utterly degrades for all the range of SNRs and the gain provided by the  DQLC scheme is even larger. %
The intuition behind this behaviour is related to the fact that \ac{MMSE} estimation for uncoded transmissions results in a weighted average of the transmitted symbols. Hence, the distortion at the receiver is influenced both by the noise variance and by the difference of the source symbols with respect to their average, which in turn depends on the source correlation. This implies that noise is the limiting factor for low SNR values, but below some noise level, the main contribution to the symbol distortion comes from the approximation in this average operation. The error caused by this strategy is constant for a given source correlation, causing the system to saturate above some SNR threshold.

Note that the obtained results also agree with the behaviour observed for DQLC mappings over \ac{AWGN} channels and show that the proposed optimization algorithm provides adequate values for the mapping parameters. They also show that the \ac{MMSE} decoder based on the idea of searching feasible combinations of quantizer intervals with the sphere decoder works correctly for the considered scenario.

Another interesting result can be observed in \Cref{fig:FigSdr_3-1x1a}. The \ac{DQLC} with a single quantized user and two uncoded ones is able to provide a slightly better performance than that of the full uncoded scheme and the 2-quantized \ac{DQLC} for a specific range of SNRs (between 20 dB  and 35 dB). As expected, the quantized transmission of one user allows to move the saturation point to a higher SNR. On the other hand, the optimization algorithm provides a similar quantization step $\Delta_1$ for the two \ac{DQLC} schemes. Hence, the gain of the one-quantized scheme is due to the fact that just in that SNR region, the linear transmission of two highly correlated symbols provides lower distortion than combining quantization and one linear transmission. In any case, the performance gain is rather small and it vanishes as the symbol correlation is lower (see \Cref{fig:FigSdr_3-1x1b}). In addition, note that as the number of users grows, it also increases the number of possible configurations for the \ac{DQLC}. For these reasons, and for simplicity, we prefer to focus on \ac{DQLC} schemes with $K_q = K-1$ for the following experiments. Finally, the performance loss caused by the use of a fixed set of parameters is remarkable for both correlation factors and in all the range of SNRs. These results highlight the importance of optimizing the mapping parameters appropriately to obtain an optimal performance of the \ac{DQLC} system.

\begin{figure}
	\centering
	\includegraphics[scale=0.9]{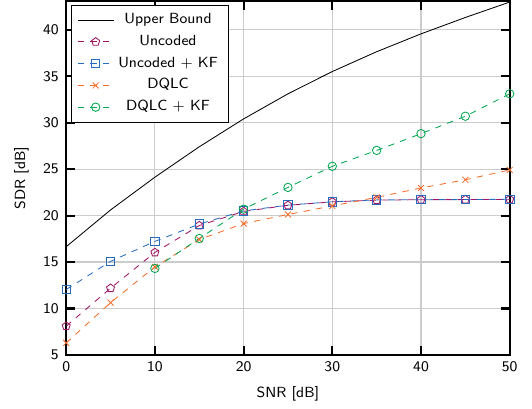}
	\includegraphics[scale=0.9]{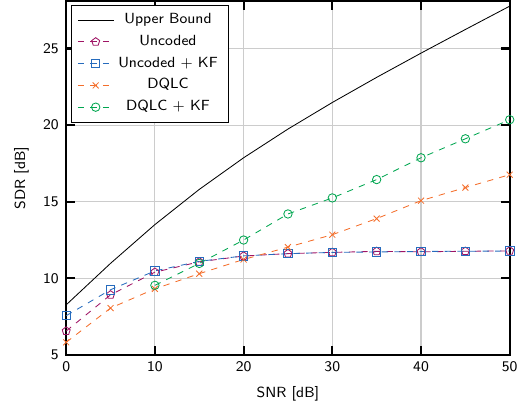}
	\caption{Performance of the different transmission schemes for $K=3$ users and two different spatio-temporal correlation factors: $\varphi=\rho =0.99$ (top) and $\varphi=\rho =0.90$ (bottom).}
	\label{fig:FigSdr_3-1x1_KF}
\end{figure}

In the next experiment, we address the case of transmitting spatial and temporally correlated sources using the DQLC scheme. A similar scenario to the previous one is considered with $K=3$ users, but now $\varphi=\rho$. In this situation, the proposed KF-based decoder is able to exploit both the temporal and the spatial correlation of the sources. \Cref{fig:FigSdr_3-1x1_KF} shows the SDR curves for four different transmission schemes: 1) \ac{DQLC} with the KF-based decoder using the prediction step a priori information; 2) \ac{DQLC} with the KF-based decoder disregarding the a priori information; 3) the linear system using the standard linear KF to decode the information; and 4) the linear system disregarding the information corresponding to the temporal correlation. In the figure,
it is also included a bound based on source-channel separation, where a lower bound on the sum rate-distortion of the multi-terminal encoding of a Gaussian variable \cite{Wang2010} is equated to the sum-capacity of the MAC channel \cite{tse2005}.
The temporal correlation is modeled by assuming a virtual system with $TK$ users and an appropriate covariance matrix to jointly represent the temporal and spatial correlation. In this case, $T$ is set to $10$ for practical reasons and because DQLC+KF systems already converge for this block size.

As observed, we consider two different correlation factors, $\rho =\varphi = 0.99$ (top) and $\rho= \varphi =0.90$ (bottom), since the benefits of the temporal correlation are more visible.  On the one hand, the improvement of using the linear KF limits to the low SNR regime since it does not provide any gain for medium and high SNRs. On the other hand, the use of the non-linear KF proposed for DQLC provides a significant performance gain, especially for high SNRs. In particular, this gain is about 9 dB for $\varphi = 0.99$ and about 3.5 dB for $\varphi = 0.90$ when the SNR is $\eta = 50$ dB. Thus, the KF-based decoder for DQLC mappings is able to raise the system performance by exploiting the temporal correlation. As expected, these gains become smaller as the temporal correlation is lower. Regarding the gap of the DQLC schemes to the upper bound, it is worth remarking that the plotted upper bound is quite optimistic for this scenario since it is computed by equating only the sum-distortion rate to the channel sum-capacity, disregarding the individual constraints on the rates. Thus, this bound will be more optimistic as the number of users increases (virtual users in this case).

\begin{figure}
	\centering
	\includegraphics[scale=1.0]{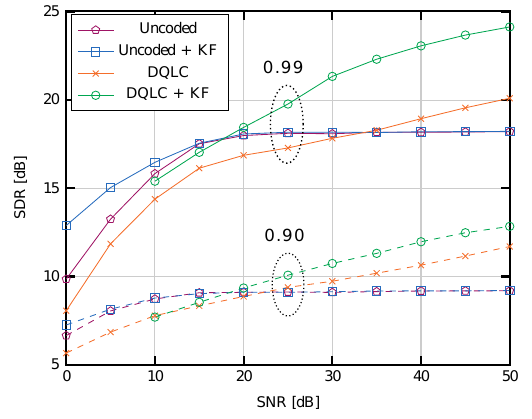}
	\caption{Performance of the different encoding schemes for $K=4$ users considering an exponential correlation model with $\rho=\varphi=0.99$ and $\rho=\varphi=0.90$.
	}
	\label{fig:Fig4Us_ExpCorr}
\end{figure}

We now explore the impact of varying the correlation model on the performance of the \ac{DQLC} schemes. For this reason, we consider an exponential model where the elements of the covariance matrix are given by $[\B{C}_{\B{s}}]_{i,j} = \rho^{|i-j|} ~~\forall i\geq j$. \Cref{fig:Fig4Us_ExpCorr} shows the obtained results for a \ac{MAC} scenario with $K=4$ users and an exponential correlation model considering two different correlation factors, $\rho=0.99$ and $\rho=0.90$. The corresponding upper bounds have been omitted in this case for clarity. As observed, we can draw similar conclusions to the previous experiment. First, the gain provided by the KF-based decoding is more remarkable as the symbol correlation increases. In addition, the gain of the simple \ac{DQLC} scheme (without KF) respect to the uncoded schemes is larger as the spatial correlation decreases and the cut point moves to lower SNRs.

\begin{figure}
	\centering
	\includegraphics[scale=1.0]{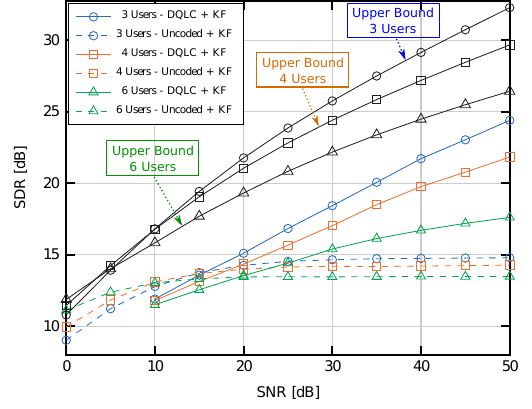}
	\caption{Performance of the DQLC and linear schemes using KF decoding for $\rho=\varphi=0.95$ and different number of users.}
	\label{fig:FigComp_UsersKF}
\end{figure}

The impact of increasing the number of \ac{MAC} users on the performance of the \ac{DQLC} system with the proposed KF-based decoder is illustrated in \Cref{fig:FigComp_UsersKF}. The SDR obtained with this scheme is compared to that of the linear system with the standard linear KF for $\rho=\varphi = 0.95$ and different number of users. The upper bounds assuming spatio-temporal correlation for each configuration are also included in the figure. As observed, the DQLC system clearly outperforms the linear scheme in the high SNR regime, although the performance gain is lower as the number of users increases: it goes from almost $10$ dB for $3$ users to over $4$ dB for $6$ users when the SNR is $\eta = 50$ dB. However, this behaviour matches to that of the standard DQLC for the \ac{AWGN} channel \cite{floor15}. Note that when the number of users is larger, it is essential to increase the parameters $\Delta_k$  to prevent the transmitted symbols cross to other intervals due to the interferences of the next users, and hence the quantization error will be larger. Finally, the gap of the DQLC+KF schemes to the corresponding upper bounds remains stable regardless of the number of users.%

\begin{figure}
	\centering
	\includegraphics[scale=1.0]{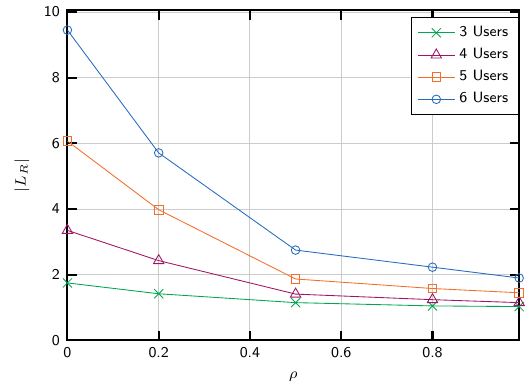}
	\caption{Number of feasible $\B{l}$ vectors provided by the sphere decoder as a function of the source correlation for $\eta=30$ dB.}
	\label{fig:FigSizeLR}
\end{figure}

Finally, \Cref{fig:FigSizeLR} shows the average size of the set of feasible $\B{l}$ vectors, $\mathcal{L}_R$, provided by the sphere decoder for different levels of source correlation and number of users, and when the SNR is $\eta=30$ dB. The radius of the sphere decoder was set according to the criterion explained in \Cref{sec:radius_sd}. As mentioned, the computational cost of the decoding operation is directly related to the size of this set because it determines the number of delimited integrals that must be solved numerically to compute the estimates for each source vector. Thus, it is important to verify if the number of elements of the set $\mathcal{L}_R$ is small for the different scenarios in order to make the decoding operation affordable. As observed, when we consider $3$ transmitters, the sphere decoder mostly selects a single candidate vector regardless of the correlation level. However, the size of the set $\mathcal{L}_R$ grows as the number of users increases and, especially, for low correlation factors. In such situations, the uncertainty in the decoding operation is larger for two reasons. On the one hand, a higher source correlation implies that the decoder will have available more a priori information that can be utilized to disregard most combinations of intervals. On the other hand, a larger number of users implies to lower the power allocated to the last users to avoid crossing effects. Hence, for a given SNR, the uncertainty due to the noise will be larger in those users. %

\section{Conclusion}

We have studied the design of practical zero-delay \ac{JSCC} schemes for fading \acp{MAC} where the source symbols are correlated both in the time and spatial dimensions. At the transmitters, the source information is encoded using a distributed scheme based on \ac{DQLC}, while a decoding approach which combines the idea of sphere decoding and non-linear KF techniques have been proposed to exploit both types of source correlation with a reasonable computational cost for an arbitrary number of users. In addition, the proposed \ac{DQLC}-based schemes can be adapted to the potential channel variations through the optimization of its parameters using an algorithm which replaces an exhaustive search on the whole parameter space, therefore with a prohibitive complexity even for a few users, by a search of the optimal power allocations for the group of users. Computer simulation results show that the proposed \ac{JSCC} scheme provides a significant performance gain with respect to uncoded transmissions for medium and high SNRs. %

\section*{Acknowledgment}

This work has been funded by Office of Naval Research Global of United States (N62909-15-1-2014), the Xunta de Galicia (ED431C 2016-045, ED341D R2016/012, ED431G/01), the Agencia Estatal de Investigación of Spain (TEC2015-69648-REDC, TEC2016-75067-C4-1-R) and ERDF funds of the EU (AEI/FEDER, UE).

\appendix

\section*{Derivation of the lattice expression for DQLC mappings}

First, we partition the covariance matrices $\B{\Sigma}_{\B{s}}$ and $\B{C}_e$ as
\begin{align}
\B{\Sigma}_{\B{s}} = \left(
\begin{array}{c|c}
\B{\Sigma}_q & \B{\Sigma}_c\\
\hline
\B{\Sigma}_c^T & \B{\Sigma}_u
\end{array}
\right)~~~~\qquad
\B{C}_e = \left(
\begin{array}{c|c}
\B{C}_q & \B{C}_c\\
\hline
\B{C}_c^T & \B{C}_u
\end{array}
\right),
\end{align}
where the subindices $q$ and $u$ refer to the part of the covariance matrices corresponding to the quantized and uncoded users, respectively, and the subindex $c$ corresponds to the part for the cross correlation. 
Next, evaluating \eqref{eq:quadratic_form} in the chosen point $\B{\tilde{s}_l}$, we start from
\begin{align}
\Omega(\B{l}, \tilde{\B{s}}_{\B{l}})=\phi_{\B{l}} \exp\left(-\frac1{2}(\tilde{\B{s}}_{\B{l}}-\B{\mu}_{\B{l}})^T\B{C}_e^{-1}(\tilde{\B{s}}_{\B{l}}-\B{\mu}_{\B{l}})\right),
\label{eq:evaluating_lattice}
\end{align}
and now we can replace $\tilde{\B{s}}_{\B{l}}$ by its expression in \eqref{eq:s_tilde}, obtaining 
\begin{align}
&\hspace*{-0.2cm} \small \Omega(\B{l}) = \notag \\
&\hspace*{-0.2cm} \small
\phi_{\B{l}} \exp
\left(-\frac1{2}\left(\B{D}\left(\B{l} + \frac{1}{2}\B{1}\right) - \B{\mu}_{\B{l}}^{(q)}\right)^T
\tilde{\B{C}}_e^{-1}\left(\B{D}\left(\B{l} + \frac{1}{2}\B{1}\right) - \B{\mu}_{\B{l}}^{(q)}\right)
\right)
\label{eq:first_lattice}
\end{align}

\hspace*{-0.35cm}with
$\tilde{\B{C}}_e = \B{\Sigma}_q - \B{\Sigma}_c^T\B{\Sigma}_u^{-1}\B{\Sigma}_c$,
and $\B{\mu}_{\B{l}}^{(q)} = [\B{\mu}_{\B{l}}]_{1:2K_q}$, the first $2K_q$ components of the linear MMSE estimates in \eqref{eq:lmmse_for_l} corresponding to the quantized users. Note that the remaining $2K-2K_q$ components in the exponent of $\Omega(\B{l}, \tilde{\B{s}}_{\B{l}})$ are not relevant since they become zero.

We next develop the left part of the exponent in \eqref{eq:first_lattice} as
\begin{align}
\B{D}\left(\B{l} + \frac{1}{2}\B{1}\right) - \B{\mu}_{\B{l}}^{(q)}= &
~\B{D}\B{l} + \frac{1}{2}\B{D}\B{1} - \bar{\B{s}}^{(q)} ~- \notag \\
&\frac{1}{\sigma_n^2}\B{C}_c \B{A}_u^T\B{H}_u^T \left(\B{y}_l - \B{H}_u\B{A}_u\bar{\B{s}}^{(q)} \right),
\end{align}
where the channel matrix was decomposed as $\B{H} = [\B{H}_q, \B{H}_u]$, $\bar{\B{s}}^{(q)}$ represents the first $2K_q$ components of the predicted mean, $\B{A}_q = \operatorname{diag}\left\{ \alpha_1,\ldots,\alpha_{K_q} \right\}$ and $\B{A}_u = \operatorname{diag}\left\{ \alpha_{K_{q}+1},\ldots,\alpha_{K} \right\}$.

Replacing $\B{y}_l$ by its expression and reordering the resulting terms, we finally obtain 
\begin{align}
&\B{D}\left(\B{l} + \frac{1}{2}\right) - \B{\mu}_{\B{l}}^{(q)} = \B{D}\B{l} + \frac{1}{2}\B{D}\B{1} - \bar{\B{s}}^{(q)} ~- \notag\\ 
&\frac{1}{\sigma_n^2}\B{C}_c \B{A}_u^T\B{H}_u^T \left(\B{y} - \B{H}_q\B{A}_q\left(\B{l}+\frac1{2}\right)  - \B{H}_u\B{A}_u\bar{\B{s}}^{(q)} \right)\notag\\
& = \B{B}\B{l} - \B{v},
\end{align}
with
\begin{align}
\B{B} &= \B{D} + \frac{1}{\sigma_n^2}\B{C}_c \B{A}_u^T\B{H}_u^T\B{H}_q\B{A}_q\\
\B{v} &= \frac{1}{\sigma_n^2}\B{C}_c \B{A}_u^T\B{H}_u^T \left(\B{y} - \B{H}_q\B{A}_q\frac1{2}\B{1} -\B{H}_u\B{A}_u\bar{\B{s}}^{(q)}\right) -
\frac{1}{2}\B{D}\B{1} +\bar{\B{s}}^{(q)}.
\end{align}
Therefore, \eqref{eq:first_lattice} can be rewritten as
\begin{align}
\phi_l \exp\left(-\frac1{2} \left( \B{B}\B{l} - \B{v} \right)^T\tilde{\B{C}}_e^{-1}\left( \B{B}\B{l} - \B{v} \right)
\right)
\label{eq:lattice1}
\end{align}
Note that the last term in \eqref{eq:lattice1} can be disregarded because it does not depend on $\B{l}$.

On the other hand, developing the expression for the weights $\phi_{\B{l}}$, we obtain 
\begin{align}
\phi_{\B{l}} &= \exp\left( -\frac1{2}\left(\frac1{\sigma_n^2}\B{y}_l^T\B{y}_l -
\frac{1}{\sigma_n^4}\B{y}_l^T\B{H}\B{G}_u\B{C}_e\B{G}_u^T\B{H}^T\B{y}_l - \right. \right. \notag \\
&\hspace*{1.3cm}\left. \left. \frac{2}{\sigma_n^2}\bar{\B{s}}^T\B{\Sigma}_{\B{s}}^{-1}\B{C}_e\B{G}_u^T\B{H}^T\B{y}_l
\right)\right) \\
 &= \exp\left(-\frac1{2}\left( \B{y}_l^T\B{Z}\B{y}_l + 2\B{u}^T\B{y}_l
\right)\right) 
\end{align} 
with
\begin{align}
\B{Z} &= \frac1{\sigma_n^2}\left(\B{I} - \frac{1}{\sigma_n^2}\B{H}\B{G}_u\B{C}_e\B{G}_u^T\B{H}^T\right) = \left(\sigma_n^2\B{I} + \B{H}\B{G}_u\B{\Sigma}_{\B{s}}\B{G}_u^T\B{H}^T\right)^{-1}\\
\B{u} &=\frac{1}{\sigma_n^2}\B{H}\B{G}_u\B{C}_e\B{\Sigma}_{\B{s}}^{-1}\bar{\B{s}}.
\end{align}

Now, replacing $\B{y}_l$ by its expression, the exponent of $\phi_{\B{l}}$ is given by
\begin{align}
&-\frac1{2}\left( \B{y}_l^T\B{Z}\B{y}_l + 2\B{u}^T\B{y}_l \right) = \notag\\
&  -\frac1{2}\left[\left(\B{y} - \B{H}_q\B{A}_q\left(\B{l}+\frac1{2}\B{1}\right)\right)^T\B{Z}~\left(\B{y} - \B{H}_q\B{A}_q\left(\B{l}+\frac1{2}\B{1}\right)\right) + \right. \notag \\
&+ \left. 2\B{u}^T\left(\B{y} - \B{H}_q\B{A}_q\left(\B{l}+\frac1{2}\B{1}\right)\right)\right] \notag
\end{align}
Reordering the resulting terms, the above expression is rewritten as
\begin{align}
& -\frac1{2}\left(\B{l}^T\B{A}_q^T\B{H}_q^T\B{Z}\B{H}_q\B{A}_q\B{l} -2\B{y}^T\B{Z}\B{H}_q\B{A}_q\B{l} ~+ \right. \notag \\
& \hspace*{0.85cm}  \left. \B{1}^T\B{A}_q^T\B{H}_q^T\B{Z}\B{H}_q\B{A}_q\B{l} - 2\B{u}^T\B{H}_q\B{A}_q\B{l} \right)=\\ &-\frac1{2}\left(\B{l}^T\B{A}_q^T\B{H}_q^T\B{Z}\B{H}_q\B{A}_q\B{l} -2\B{m}^T\B{l}
\right),
\label{eq:lattice2}
\end{align}
with 
\begin{align}
\B{m} = \B{A}_q^T\B{H}_q^T\left(\frac{1}{\sigma_n^2}\B{H}\B{G}_u\B{C}_e\B{\Sigma}_{\B{s}}^{-1}\bar{\B{s}} + \B{Z}\left(\B{y} -\frac1{2}\B{H}_q\B{A}_q\B{1}\right)\right).
\end{align}

Combining the exponents in \eqref{eq:lattice1} and \eqref{eq:lattice2} into a single exponent, we finally obtain
\begin{align}
\Omega(\B{l}) & = \exp\left(-\frac1{2}\left[\B{l}^T\left(\B{B}^T \tilde{\B{C}}_e^{-1}\B{B} + \B{A}_q^T\B{H}_q^T\B{Z}\B{H}_q\B{A}_q\right)\B{l} ~- \right. \right. \notag \\
& \hspace*{1.9cm}\left. \left. 2\left(\B{v}^T \tilde{\B{C}}_e^{-1}\B{B} + \B{m}^T\right)\B{l}\right]\right).
\end{align}

The last step consists in expressing the above exponent in a lattice form such as
\begin{align}
\Omega(\B{l}) = \exp\left(-\frac1{2}\left(\B{l} - \B{l}_o \right)^T\B{\Lambda}\left( \B{l} - \B{l}_o \right)\right), 
\end{align}
where
\begin{align}
\B{\Lambda} &= \B{B}^T \tilde{\B{C}}_e^{-1}\B{B} + \B{A}_q^T\B{H}_q^T\B{Z}\B{H}_q\B{A}_q,\\
\B{l}_o &= \Lambda^{-1}\left(\B{B}^T\tilde{\B{C}}_e^{-1}\B{v} + \B{m}\right).
\end{align}

\bibliographystyle{IEEEtranTCOM}
\bibliography{references}

\end{document}